\documentclass[12pt]{emulateapj}
\usepackage{color}

\newcommand\ha{H$\alpha$~}

\shorttitle{Nature of \ha selected galaxies . II.}
\shortauthors{Tadaki et al.}

\begin{document}


%\title{Nature of \ha selected galaxies at $z>2$. I. Dusty star formation of main sequence galaxies}
\title{Nature of \ha selected galaxies at $z>2$. II. clumpy galaxies and compact star-forming galaxies}


\author{Ken-ichi Tadaki\altaffilmark{1}, Tadayuki Kodama\altaffilmark{1,2}, Ichi Tanaka\altaffilmark{3}, Masao Hayashi\altaffilmark{4}, Yusei Koyama\altaffilmark{1}}
\email{tadaki.ken@nao.ac.jp}

\author{Rhythm Shimakawa\altaffilmark{2}}


\altaffiltext{1}{Optical and Infrared Astronomy Division, National Astronomical Observatory of Japan, Mitaka, Tokyo 181-8588, Japan}
\altaffiltext{2}{Department of Astronomical Science, The Graduate University for Advanced Studies, Mitaka, Tokyo 181-8588, Japan}
\altaffiltext{3}{Subaru Telescope, National Astronomical Observatory of Japan, 650 North A'ohoku Place, Hilo, HI 96720, USA}
\altaffiltext{4}{Institute for Cosmic Ray Research, The University of Tokyo, 5-1-5 Kashiwanoha, Kashiwa, Chiba 277-8582, Japan}


% measure sizes and S$\acute{\mathrm{e}}$rsic indices of the HAEs 
%Most of the HAEs have disk-like morphologies and their size-stellar mass relation is similar to local one, supporting the universal relation reported by previous studies. 

\begin{abstract}
We present the morphological properties of 109 H$\alpha$-selected galaxies at $z>2$ in SXDF-UDS-CANDELS field. 
With high-resolution optical/near-infrared images obtained by $Hubble\ Space\ Telescope$, we identify giant clumps within the \ha emitters (HAEs). 
We find that at least 41\% of our sample show clumpy structures in the underlying disks.
The color gradient of clumps is commonly seen in the sense that the clumps near the galactic center tend to be redder than those in the outer regions.
The mid-infrared detection in galaxies with red clumps and the spatial distribution of \ha emission suggest that dusty star-formation
activity is probably occurring in the nuclear red clumps.
A gas supply to a bulge component through the clump migration is one of the most potent physical processes to produce such dusty star-forming clumps and form massive bulges in local early-type galaxies.
They would become large quiescent galaxies at later times just by consumption or blowout of remaining gas.
Also, while most of the HAEs have extended disks, we observe two massive, compact HAEs, whose stellar surface densities are significantly higher. 
They are likely to be the direct progenitors of massive, compact quiescent galaxies at $z=1.5$--2.0.
Two evolutionary paths to massive quiescent galaxies are devised to account for both the size growth of quiescent galaxies and their increased number density from $z\sim2$ to $z=0$.
\end{abstract}


\keywords{galaxies: evolution -- galaxies: high-redshift}



\section{Introduction}

By the recent advances on the multi-wavelength observing techniques,
intensive quantitative studies have been conducted on global properties of high-redshift galaxies, such as stellar mass, SFR, age and metallicity of stellar populations.
An important next step is to spatially resolve their properties and 
identify the physical processes that govern the galaxy formation and evolution.
However, it is difficult to dissect distant galaxies with ground-based telescopes as the spatial resolution of 1\arcsec\ corresponds to $\sim$8 kpc in the physical scale at $z\simeq2$.
The high-resolution imaging with $Hubble\ Space\ Telescope$ (HST) allows us to study the evolution of galaxy morphologies in the distant universe.
The Advanced Camera for Surveys (ACS) on HST has demonstrated that star-forming galaxies at $z>2$ tend to have irregular, clumpy morphologies unlike the present-day galaxies on the Hubble sequence \citep[e.g.,][]{2005ApJ...627..632E,2006ApJ...636..592L,2007ApJ...656....1L}.
However, the optical camera ACS can sample only the rest-frame ultraviolet (UV) lights from high-redshift galaxies, and its images pick out only star-forming regions within galaxies rather than the bulk structures of galaxies traced by stellar mass.
With the advent of Near Infrared Camera and Multi-Object Spectrometer (NICMOS) and Wide Field Camera 3 (WFC3), it has become possible to measure the rest-frame optical morphologies for statistical samples of galaxies at $z>2$.
Surprisingly, clumpy structures are often seen in both rest-frame UV and optical wavelengths, especially for massive star-forming galaxies \citep[e.g.,][]{2009ApJ...692...12E,2011ApJ...731...65F,2012ApJ...753..114W}.
Clumps are characterized typically by a large size of $\sim$1 kpc and a large stellar mass of $\sim10^9M_\odot$ for luminous star-forming galaxies with log $F_{\mathrm{obs,H}\alpha} [\mathrm{erg\ s}^{-1}] >42.6$ \citep{2011ApJ...739...45F}.
Although such clumpy structures are first interpreted as the outcome of a major merger \citep{2003ApJ...596L...5C}, recent dynamical studies of high-redshift galaxies suggest an alternative origin.

Spectroscopies with an AO-assisted integral field unit (IFU) on 8--10m class telescopes have enabled us to study in detail the internal kinematics of star-forming galaxies at $z\sim2$ as well as the spatial structures \citep[e.g.,][]{2006Natur.442..786G,2009ApJ...697.2057L,2012MNRAS.426..935S}.
To date, the largest survey with near-infrared IFU spectroscopy for galaxies at $z=1-3$ is the work by \cite{2009ApJ...706.1364F}.
They have obtained the \ha spectra of 50 star-forming galaxies at $z=1.3-2.6$, selected in the rest-frame UV, optical, near-infrared and submillimeter, with SINFONI on the VLT.
Although their morphologies in the \ha line emission are generally irregular or clumpy like their appearances in the rest-frame UV/optical continua images,
their kinematics of ionized gas show ordinary disks with symmetric rotation of $v_\mathrm{rot}\sim200-300$ km s$^{-1}$.
Many of them also exhibit a large local velocity dispersion of $\sigma=30-90$ km s$^{-1}$, suggesting that the gas disks commonly have random motions.
About one third of their \ha IFU sample are classified as rotation-dominated galaxies with $v_\mathrm{rot}/\sigma>1$. 
Roughly another third are categorized into dispersion-dominated galaxies.
These kinematical classifications strongly correlate with their intrinsic sizes in the sense that smaller galaxies are more likely dispersion-dominated \citep{2013ApJ...767..104N}.
%This implies that dispersion-dominated galaxies actually have the intrinsic properties similar to rotation dominated galaxies, but they are primarily more compact and have lower masses corresponding to an earlier stage of evolution.
Although a lot of clumpy star-forming galaxies have been observed at high redshift, interacting or merging systems inferred from dynamical signatures are only about one third \citep{2011ApJ...739...45F}.
However, we should be cautious of the face value because it is strongly biased by the sample selection of the IFU observations, which tend to pick up a bright \ha emission.
Since a merger system is likely to be associated with a bright \ha line emission,
the intrinsic merger fraction can be smaller than that.

While $z\sim2$ is the epoch when cosmic star-formation activities in galaxies come to a peak \citep{2006ApJ...651..142H}, 
some massive quiescent galaxies already exist at the same
epoch \citep[e.g.,][]{2006ApJ...649L..71K,2009ApJ...700..221K}.
It is remarkable that these quiescent galaxies are extremely compact with an effective radius of only $\sim$1 kpc \citep[e.g.,][]{2006MNRAS.373L..36T,2007ApJ...671..285T,2008ApJ...677L...5V}.
This is about a factor of four to five smaller than the size of local giant ellipticals at the fixed stellar mass.
In the local Universe, such compact, red sequence galaxies are extremely rare \citep{2010ApJ...720..723T}.
These compact quiescent galaxies must have evolved significantly in size after $z\sim2$ much faster than in mass, hence by mechanisms other than simple star formation.
The most widely accepted scenario at the moment is the inside-out growth through repeated minor mergers and accretion of small satellites \citep[e.g.,][]{2009ApJ...699L.178N,2013MNRAS.428..641O}.
Dissipationless minor mergers are more effective to enlarge the compact galaxies in size by putting mass efficiently to the outer region of galaxies.
In contrast, major mergers lead to growth both in size and mass proportionally, and are not supported by the local relation between $r_\mathrm{e}$ and $M_*$ \citep{2009ApJ...697.1290B}.
\cite{2010ApJ...709.1018V} have investigated the redshift evolution of mass profiles based on the samples selected at the constant number density in co-moving space so that the direct comparison can be made.
They have shown that the mass within the radius of 5 kpc stays nearly constant with redshift, whereas the mass at 5 kpc $< r <$ 75 kpc has increased by a factor of four from $z = 2$ to zero.
\cite{2013ApJ...766...15P} have improved this study of mass profile evolution with HST images of high spatial resolution, and 
confirm that there is no substantial growth at $z < 2$ in the central part of $r <$ 2 kpc.

The merger paradigm accounts for both the enhancement of SFR (e.g.,\ increase in starburst galaxies such as SMGs) at high redshift and the transformation of galaxy morphologies.
However, the dominant populations responsible for the cosmic stellar mass density and SFR density are found not to be merger-driven starburst galaxies but normal star-forming ($main$ $sequence$) galaxies \citep{2011ApJ...739L..40R}.
An alternative process is required to account for enhancement in SFRs, clumpy structures embedded in ordinary rotational disks, large velocity dispersions and compact sizes.
In the past several years, the efficient gas supply through cold streams along filamentary structures is suggested to be a preferred mechanism to account for these properties (clumpy structure and compact system) of high-redshift galaxies \citep{2009Natur.457..451D}.
The steady, narrow, cold gas streams penetrate the shock-heated media of massive dark matter halos and continuously supply a large amount of gas to the inner regions of star-forming galaxies.
A cosmological simulation implies that there are two types of streams:\ smooth mode and clumpy mode \citep{2009ApJ...703..785D}.
In massive halos with $M_\mathrm{halo}\sim10^{12} M_\odot$ at $z>1$, smooth streams maintain dense gas-rich disks, as observed in CO observations \citep{2010Natur.463..781T,2010ApJ...713..686D}.
In such a gas-rich disk, a small gas perturbation in the radial direction would grow up and fragment into giant clumps by gravitational instability.
Then, the internal gravitational interaction within a perturbed disk generates turbulence and the velocity dispersion becomes progressively larger. 
In fact, theoretical models provide support for gas-rich turbulent disks having clumps \citep{2007ApJ...670..237B, 2009ApJ...694L.158B}.
On the other hand, the clumpy components in cold streams lift up the velocity dispersion in disks.
The external clumps merge into the central spheroid as a wet major merger and form a compact spheroid that is significantly smaller in size than the disk \citep{2006MNRAS.370.1445D}.
Though no direct observational evidence for cold gas streams has been found yet,
these two kinds of gas stream may account for the differences in the properties between high redshift and local galaxies.

Recent large surveys have dramatically improved our understanding of the physical properties of star-forming and quiescent galaxies at $z>2$.
However, not much is yet known about the evolutionary paths from star-forming galaxies to quiescent galaxies.
Remaining critical issues include: What fraction of star-forming galaxies is in the clumpy phase? 
What are their fates? 
How are they connected to compact or extended quiescent galaxies at later times, and how does that transition happen?
To shed light on these issues, what is required is to study galaxy morphologies with a ``clean,'' ``statistical'' and ``unbiased'' sample of star-forming galaxies at this critical era of galaxy formation at $z\sim2$ and beyond, and then to spatially resolve their internal properties within galaxies such as clumpy structures, extended disks, or compact bulges.
Many previous studies had to rely on the sample selection by photometric redshifts.
Due to the significantly degraded accuracy of photometric redshifts at $z>2$,
however, we are left with incomplete, biased samples of star-forming galaxies at this epoch.

In this paper, we study the morphologies of \ha emitters (HAEs)
at $z>2$, which is a robust star-forming galaxy sample selected by narrow-band (NB) imaging.
This paper is structured as follows: In Section 2, we note the existing data and our sample in Subaru/$XMM$--$Newton$ Deep survey Field (SXDF). 
The analyses to characterize the morphological properties of HAEs and their results are described in Section 3.
We show the clump properties within HAEs and newly discovered compact star-forming galaxies at $z>2$ in Section 4.
In Section 5, we discuss the evolutionary track from HAEs at $z>2$ to massive, quiescent galaxies in the present Universe.
We summarize our study in Section 6.
Throughout this paper, we assume the cosmological parameters of H$_0$=70 km s$^{-1}$ Mpc$^{-1}$, $\Omega_M=0.3$, and $\Omega_\Lambda=0.7$, and Salpeter IMF is adopted for the estimation of stellar masses and SFRs \citep{1955ApJ...121..161S}.

\section{Sample and data}
\label{sec;sample}

\begin{figure*}
\begin{center}
\includegraphics[scale=0.85]{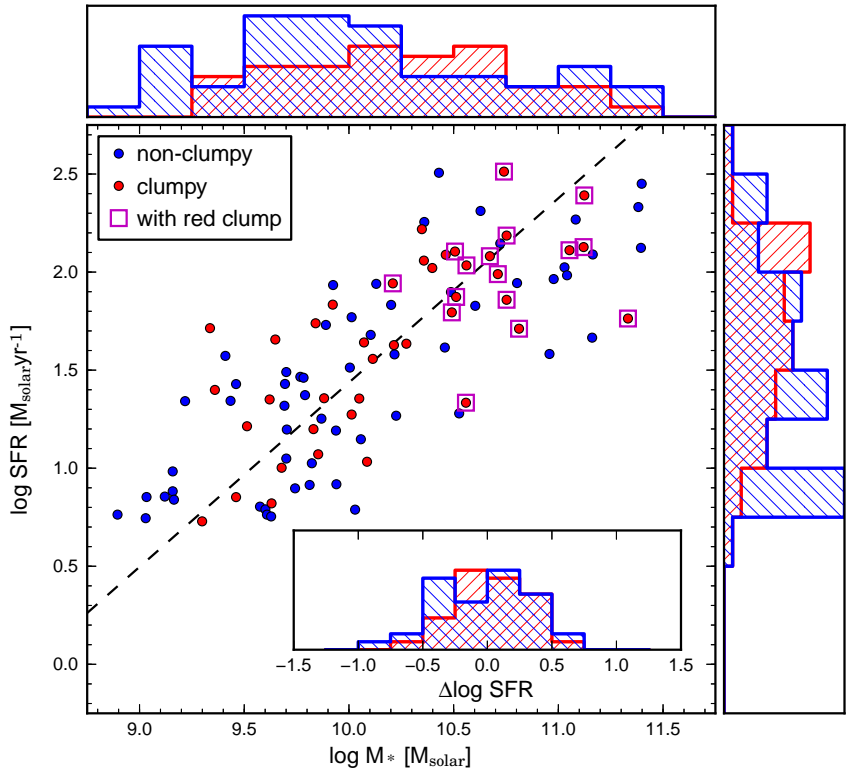}
\caption{Star-formation rates are plotted against stellar masses for the HAEs at $z=2.2$ and 2.5. Red and blue circles show clumpy and non-clumpy galaxies, respectively. A dashed black line denotes the fitted main sequence of SFR=238($M_*/10^{11}M_\odot)^{0.94}$. The histograms in the top and right panels show the projected distributions of $M_*$ and SFR for clumpy (red) and non-clumpy (blue) galaxies, respectively. In the inset of the lower corner of the main panel, we show histograms of the deviations from the main sequence in the SFR direction. Magenta squares show the objects with a red clump of $I_{814}-H_{160}>1.5$ (Section \ref{sec;clump_migration}). \label{fig;MS_clump}}
%The color images were created from $V_{606}I_{814}H_{160}$-band maps.
%Yellow solid and dashed line denotes the fitted main sequence of SFR=234($M_*/10^{11}M_\odot)^{0.93}$ and the locus 4 times above the main sequence, respectively. 
\end{center}
\end{figure*}

We use data from the ``$MAHALO$-$Subaru$'' project (MApping HAlpha and Lines of Oxygen with Subaru; see an overview by \citealt{2013IAUS..295...74K}),
This project provides a statistical analysis of \ha emitting star-forming galaxies over the peak epoch of galaxy formation ($z=1-3$), and across various environments from blank fields \citep{2011PASJ...63S.437T} to clusters/proto-cluster regions \citep{2013MNRAS.428.1551K,2011PASJ...63S.415T,2012ApJ...757...15H}.
In this paper, we use 63 HAEs at $z=2.19\pm0.02$ and 46 at $z=2.53\pm0.02$ in the SXDF-UDS-CANDELS blank field, which have been identified with two NB filters.
The observations, sample selection and estimate of physical properties are described in detail in the companion paper (Tadaki et al., in preparation).
This is a clean sample of star-forming galaxies, and importantly, much less biased to a certain limit in SFRs.
Actually, in our spectroscopic follow-up observations, 12 HAEs out of 13 targets ($\sim92$\%) have been confirmed to be located at $z=2.19$.
Five AGN-dominated objects at $z=2.19$ are rejected from the sample, based on the [N~{\sc ii}]/H$\alpha$ line ratios, IRAC colors (Tadaki et al., in preparation) and/or [C~{\sc iv}] line detections \citep{2012MNRAS.421.3060S}.
Since most of the other HAEs are located in the $blue$ $cloud$ on the color-magnitude diagram, they are expected to be star-forming galaxies rather than AGN-dominated objects (Tadaki et al., in preparation).
Their stellar masses and amounts of dust extinction are estimated from the spectral energy distributions (SEDs) with 12-band photometries from $u$-band to 4.5 $\mu$m, which are fitted by the stellar population synthesis model of \cite{2003MNRAS.344.1000B}.
The SFRs are computed from the \ha luminosities with the standard calibrations of \cite{1998ARA&A..36..189K}.
In Figure \ref{fig;MS_clump}, we show the SFR versus stellar mass diagram of our HAE sample.
The sample includes the $main$ $sequence$ galaxies with $M_*=10^{9-11.5} M_\odot$ and SFR=5--400$M_\odot$yr$^{-1}$.

To investigate the morphologies of our HAEs,
we make use of the publicly available, high spatial resolution images
at four passbands ($V_{606}, I_{814}, J_{125}$ and $H_{160}$) of the CANDELS survey \citep{2011ApJS..197...35G,2011ApJS..197...36K}.
The spatial resolution of all the images is matched to FWHM=0.18\arcsec, which is the PSF size of the $H_{160}$ image.
At $z=2$, this corresponds to 1.5 kpc in the physical scale.
The wavelength regime of the ACS-$V_{606}$ image corresponds to the rest-frame UV at $z>2$
and hence approximately traces internal star-formation activities of galaxies (with some
complexity introduced by dust extinction though).
The WFC3-$H_{160}$ image corresponds to the rest-frame optical at $z>2$
and hence approximately traces the stellar mass distribution.
As such, the ACS and WFC3 data are both imperative to investigate the morphologies of high redshift star-forming galaxies.

\begin{figure*}
\begin{center}
\includegraphics[scale=0.9]{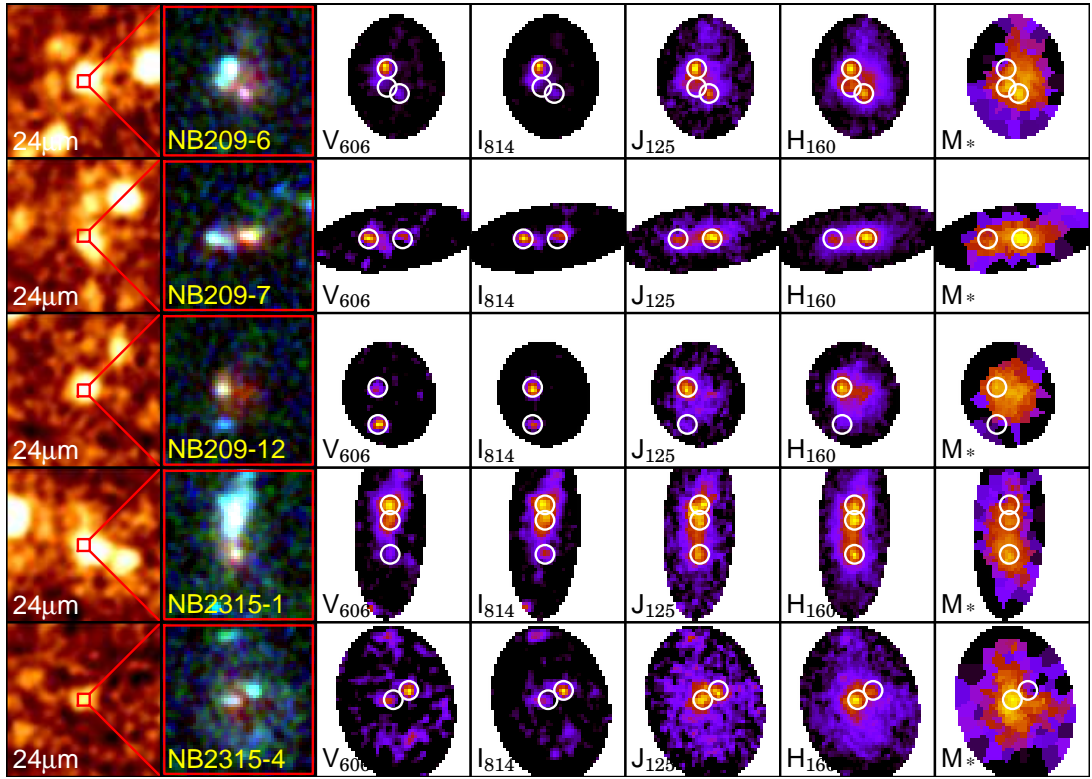}
\caption{Five examples of our clumpy HAEs. From left to right, MIPS 24$\mu$m images, three color images of $V_{606}I_{814}H_{160}$, $V_{606}$-band, $I_{814}$-band images by ACS, $J_{125}$-band, $H_{160}$-band images by WFC3, and the estimated spatial distribution of stellar mass
(see text for the estimation technique).
%%The stellar mass is calculated with the equation (2). See text for details.
The size of each image is 3\arcsec on each side, except for the MIPS images (40\arcsec$\times$40\arcsec).
Circles indicate the position of identified clumps. \label{fig;clumpy_example}}
\end{center}
\end{figure*}

\section{Analysis}

\subsection{Clump identification}
\label{sec;identification_clump}

Star-forming galaxies are found to become increasingly irregular at higher redshift ($z>2$) with clumpy structures and asymmetry \citep[e.g.,][]{2005ApJ...627..632E}.
Such clumpy nature must be mirroring important physics of the early phase of galaxy formation.
To investigate the properties of clumpy galaxies, we first identify clumps in the HAEs by a semi-automatic method using the 2-D version of $clumpfind$ code \citep{1994ApJ...428..693W} as employed by \cite{2012MNRAS.427..688L}.
Clumps are defined in the $H_{160}$ or $V_{606}$ images.
Note that $V_{606}$-band traces the rest-frame UV, dominated by massive stars,
and it gives a biased view of the clumps rather than overall stellar populations.
We thus use $I_{814} - H_{160}$ as color of clumps. (section \ref{sec;stellar_mass} and \ref{sec;clump_migration}).
Because the ACS images are not available for four HAEs at $z=2.2$, we use a total of 100 galaxies.
To minimize misidentification, the contour intervals of 3$\sigma$ are adopted.
To distinguish clumps from noises, we only accept the clumps that satisfy the following criteria : (1) the size is larger than the PSF (0.18\arcsec), and (2) the peak flux density is above the 10$\sigma $ level in $H_{160}$ map or the 6$\sigma $ level in $V_{606}$ map.
We adopt different criteria for the $H_{160}$ map and the $V_{606}$ map to match the visual inspection.
The stellar mass of identified clumps roughly corresponds to $M_\mathrm{clump}\sim10^9~M_\odot$ (Section \ref{sec;stellar_mass}).

Figure \ref{fig;clumpy_example} displays some examples of the color images of clumpy galaxies.
The identified clumps are marked by black circles.
Here, a clumpy galaxy is defined as an object consisting of multi-components, no matter whether a second-component (or a later component) is a clump formed by a gravitational collapse or an external small galaxy.
We find that 41 out of 100 HAEs ($\sim41$\%) have sub-clumps along with the main-component.
Such clumpy galaxies are seen everywhere on and around the main sequence over a wide range in stellar mass, but the fraction of clumpy galaxies peaks in $M_*\sim10^{10.5}M_\odot$ and SFR $\sim100~M_\odot$yr$^{-1}$ (Figure \ref{fig;MS_clump}).
%they tend to have slightly higher stellar mass and higher SFRs
%The fraction of clumpy galaxies is different by stellar mass of host galaxies; 33\% in $M_*>10^{11}M_\odot$, 59\% in $M_*=10^{10-11}M_\odot$ and 32\% in $M_*<10^{10}M_\odot$.
%This difference may be closely related to the effect by spatial resolution.
The fraction of clumpy galaxies noted here is the lower limit because non-clumpy galaxies could consist of a few smaller clumps, which are not resolved in the 0.18\arcsec resolution images, especially for less massive galaxies.
On the other hand, the fraction seems to decrease in the most massive galaxies with $M_*>10^{11}~M_\odot$, although clumps, if they exist, could be more easily resolved in those larger galaxies.
If clumps are continuously formed by gravitational collapses, their host disks should be gas rich so as to be able to collapse by overcoming a large velocity dispersion and a shear of differential rotation.
Given that a gas supply through a cold stream is prevented by a virial shock in a massive halo, our HAEs with $M_*>10^{11}~M_\odot$ would exhaust the bulk of gas and become unable to form new clumps.
In fact, the sSFRs of massive HAEs are suppressed compared to other HAEs, suggesting that they may be just quenching their star-formation activities.

Also, we investigate the fraction of clumpy galaxies for the HAEs with $10^{9.5}~M_\odot<M_*<10^{10.8}~M_\odot$ as a function of the offset from the main sequence at fixed stellar mass: $\Delta$log(SFR)=log(SFR)-log(SFR$_{\mathrm{MS}}$). 
The AGN fraction is likely to be small in this mass range (Tadaki et al., in preparation).
Here, the main sequence is defined as SFR$_\mathrm{MS}$=238($M_*/10^{11}M_\odot)^{0.94}$ (Tadaki et al., in preparation).
However, there is no clear evidence for a difference between clumpy and non-clumpy galaxies on the stellar mass--SFR diagram.
To identify any systematic difference in the physical properties between clumpy/non-clumpy galaxies,
we would need to investigate other quantities such as the local velocity dispersion.
In this paper, we simply conclude that the clumpy signature is common among star-forming galaxies at $z>2$.

%The Kolmogorov-Smirnov (K-S) test suggests that the null hypothesis, that they are drawn from the same distribution, is proven to be false at significance level of 4.7 \% ($\sim2\sigma$).

\subsection{Distribution of stellar components}
\label{sec;stellar_mass}

To derive the spatial distribution of stellar components within a galaxy with high precision,
we take into account the mass-to-light ratio at each position within the galaxy.
We therefore first establish a relation between the $I_{814}-H_{160}$ color and the stellar
mass-to-light ratio at $H_{160}$ band by using the population synthesized bulge-disk composite
model of \cite{1999MNRAS.302..152K}.
We note that the relation between color and mass-to-light ratio does not depend much
on the assumed star-formation histories or metallicity or dust extinction because of
the age--metallicity--dust-extinction degeneracy \citep{2003MNRAS.346....1K} as far as
the IMF is fixed.
The stellar mass at each position within galaxies is then calculated from
the $H_{160}$-band luminosity and the mass-to-light ratio derived from the
$I_{814}-H_{160}$ color. We used the following conversion equations:

\begin{eqnarray}
\mathrm{log}(M_*/10^{11})&=&-0.4(H_{160}-H_{11})\\
\Delta \mathrm{log}M_*&=&a\exp [b(I_{814}-H_{160})],
\end{eqnarray}

\noindent
where $H_{11}$=22.8, $a=-2.3$ and $b=-1.4$ at $z=2.2$, and $H_{11}$=23.0, $a=-2.4$ and $b=-1.5$ at $z=2.5$.
The equation (1) exhibits the stellar mass--magnitude relation in the case of
passively evolving galaxies at each redshift.
The other equation (2) gives the amount of correction in stellar mass as a function of
$I_{814}-H_{160}$ colors, which represents the variation in mass-to-light ratio depending on
star-formation history. 
Because the color measurement is less accurate at the pixels with low surface brightness,
we perform an adaptive spatial binning with the Voronoi technique \citep{2003MNRAS.342..345C}
so as to achieve the minimum S/N level of 10 per bin on the $H_{160}$-band image.

We present the stellar mass maps of some clumpy HAEs along with the HST images in Figure \ref{fig;clumpy_example}.
In the inferred stellar mass distribution, some bright clumps in the $V_{606}$-band become
less prominent while some other bright clumps in the $H_{160}$-band
become more prominent.
This suggests that the bright clumps seen in the $V_{606}$-band are young and star-forming,
and thus, do not contribute significantly to the overall mass distributions of the galaxies.
Moreover, it should be noted that a diffuse component is now seen in the $H_{160}$-band image or the stellar mass distribution just surrounding the geometric center of the galaxy.
Whereas these galaxies may simply resemble a merging system in the $V_{606}$-band image, 
the stellar mass is very smoothly distributed, and bridging the distinct components seen in the $V_{606}$-band image.
In the case of a merging galaxy, we would have to see multi-components in the stellar mass distribution.
Such smooth distributions of stellar components in high redshift star-forming galaxies are also supported by the analysis of spatially resolved SEDs \citep{2012ApJ...753..114W}.

We cannot clearly tell with the current existing data alone whether this is a clumpy galaxy or a merger system.
However, some recent IFU studies have been just resolving such internal kinematical
structures in some clumpy star-forming galaxies at $z\sim2$, and revealing that they often have disks with coherent rotation \citep[e.g.,][]{2006Natur.442..786G,2009ApJ...706.1364F}.
Therefore our sample of clumpy galaxies are also more likely single galaxies with internal large gravitational fragmentations of the gas in the disks at the early phase of galaxy
formation fed by abundant gas accretion \citep{2011ApJ...733..101G}. 

\subsection{Structural parameters}

\begin{figure}
\begin{center}
\includegraphics[scale=0.7]{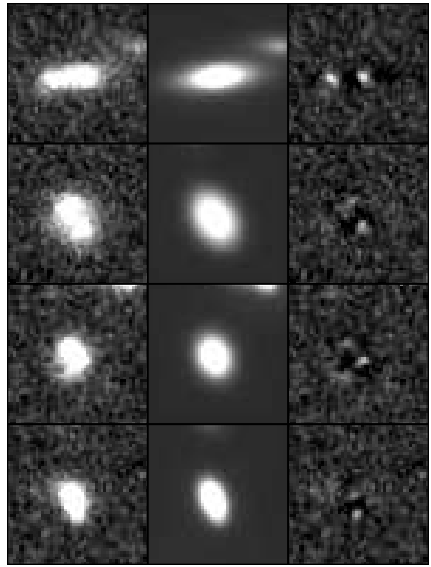}
\caption{Examples of the GALFIT model to four HAEs. $H_\mathrm{160}-$band image (left), models (center) and residuals (right) are presented for each galaxy. These galaxies are all reasonably well fitted with the S$\acute{\mathrm{e}}$rsic model.\label{fig;GALFIT}}
\end{center}
\end{figure}

The high-resolution images by WFC3 enable us to estimate the structural parameters such as size and radial profile in the rest-frame optical for galaxies at $z=2-2.5$.
The publicly available catalog by \cite{2012ApJS..203...24V} is used to obtain the structural parameters of our HAEs.
We give a brief summary of the catalog and their work below.
For all the objects detected in the CANDELS fields, the surface brightness distributions in the $H_{160}$-band images are fitted with S$\acute{\mathrm{e}}$rsic model \citep{1968adga.book.....S} by using the GALFIT version 3.0 code \citep{2010AJ....139.2097P} and GALAPAGOS \citep{2012MNRAS.422..449B}.
The free parameters are the position of galaxy center, S$\acute{\mathrm{e}}$rsic index, effective radius, axial ratio, position angle of the major axis, and total magnitude. 
Initial guesses for these parameters are taken from the SExtractor measurements.
%The fitting ranges are constrained to keep the S$\acute{\mathrm{e}}$rsic index between 0.2 and 8, effective radius between 0.02 and 24 arcsec, axis ratio between 0.0001 and 1, and magnitude between 0 and 40 as well as between $-$3 and +3 magnitudes from the input values.
Simulations with artificial objects can estimate the systematic uncertainties in the measured parameters by comparing them to the input (true) parameters.
In the case of galaxies with $H_{160}<24$, the measurement errors are estimated to be $\Delta n=-0.01 \pm 0.24$ and $\Delta r_e=-0.01 \pm 0.08$ for $n<$3, and $\Delta n=-0.25 \pm 0.33$ and $\Delta r_e=-0.22 \pm 0.19$ for $n>$3 \citep{2012ApJS..203...24V}.
Because most of the faint objects with $H_{160}>24$ are less massive galaxies than $M_*<4\times10^{9}M_\odot$, we apply the stellar mass cut of $M_*>4\times10^{9}M_\odot$ in the analysis.
However, even the sample of 86 HAEs that satisfy this criterion includes 13 faint objects.
We do not use them because the accuracy of the fitting is significantly degraded, especially for bulge-dominated galaxies.
We also reject four objects whose fitting results are bad and unreliable (flag value of two in \citealt{2012ApJS..203...24V} catalog).
After all, we have obtained the structural parameters for 69 HAEs with $M_*>4\times10^{9}M_\odot$.

\begin{figure*}
\begin{center}
\includegraphics[scale=1.0]{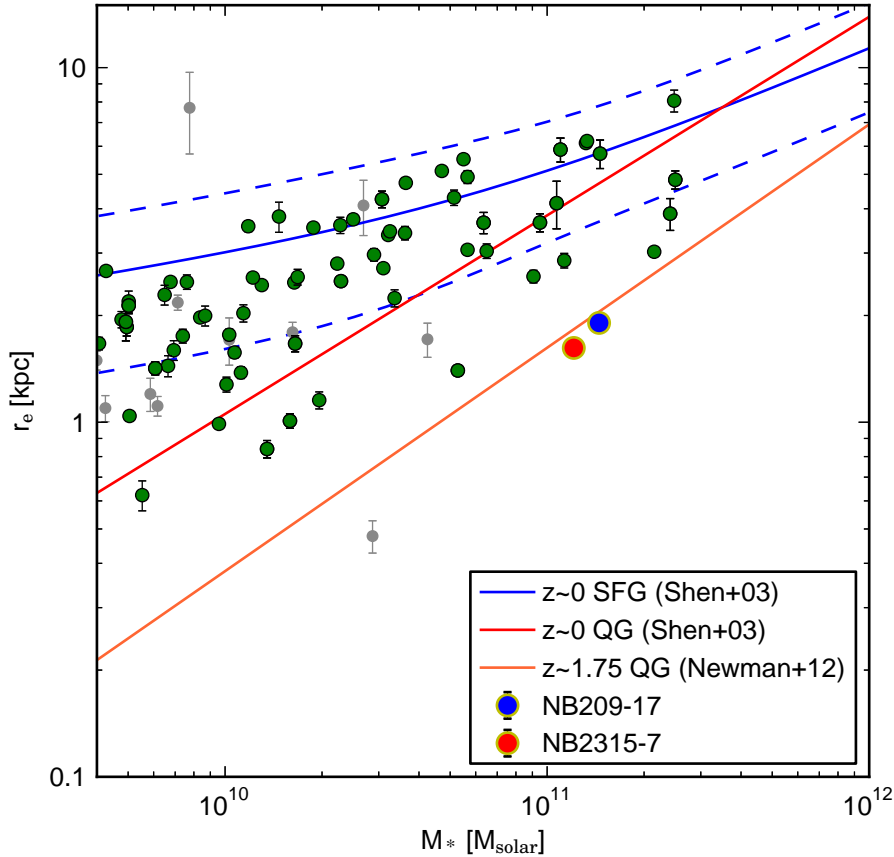}
\caption{Stellar mass--size relation of the HAEs at $z$=2.2 and 2.5 with $H_{160}<24$ (green circles), and with $H_{160}=24-26$ (gray points). The blue solid and dashed curves indicate the local relation for late-type galaxies and its 1$\sigma$ scatter, respectively \citep{2003MNRAS.343..978S}. The red solid line shows the local relation for early-type galaxies. The orange line represents the typical location of the compact quiescent galaxies at $z=1.5-2.0$ \citep{2012ApJ...746..162N}. Blue and red circles exhibit the two compact HAEs, SXDF-NB209-17 and SXDF-NB2315-7 (Section \ref{sec;progenitor_quiescent}). \label{fig;M_Re}}
\end{center}
\end{figure*}

In Figure \ref{fig;GALFIT}, we show the $H_{160}$-band images, model and residual images of
four examples of the HAEs.
These are all HAEs with star-forming clumps as identified in Section \ref{sec;identification_clump}, but the surface brightness distributions of overall galaxies seem to be suitably fitted by the S$\acute{\mathrm{e}}$rsic model.
Therefore, we do not discern between clumpy and non-clumpy galaxies when discussing their size measurements.

Seventy-four percent of them are found to have disk-like morphologies where their S$\acute{\mathrm{e}}$rsic indices lie close to unity ($n<2$). 
The median value of $n$ for the HAEs is $n=1.2$.
Other studies have also shown that high-redshift star-forming galaxies tend to have disk-like
morphologies with $n\simeq1$ \citep[e.g.,][]{2012ApJ...761...19Y,2013MNRAS.432.2012T}.
The other fourth part of our sample exhibits a high S$\acute{\mathrm{e}}$rsic index of $n>$2 indicative of bulge-dominated morphologies.

Figure \ref{fig;M_Re} shows the stellar mass--size relation of our HAEs at $z$=2.2 and 2.5,
compared to the local relation \citep{2003MNRAS.343..978S}. % and to the massive quiescent galaxies at $z\sim2$ \citep{2008ApJ...677L...5V}. 
This diagram is a powerful tool to investigate the size evolution of galaxies.
We fit the mass-size relation of log $r_\mathrm{e}$=$\gamma+\beta(\log M_*-11)$ to our HAEs, that gives $\gamma=0.58$ and $\beta=0.28$.
Most of the HAEs with $M_*>10^{10} M_\odot$ are distributed within the 1$\sigma$ range of the local relation for late-type galaxies although the star-formation activities of our HAEs are significantly higher (SFR=5--400 $M_\odot$yr$^{-1}$) than those of local disk galaxies.
This trend is consistent with the universal relation, which does not show a size evolution at a given mass, as reported by \cite{2012MNRAS.422.1014I} and \cite{2013MNRAS.430.1158S}.
On the other hand, less massive HAEs are slightly smaller than the local relation.
Even if faint galaxies with $H_{160}=24$--26 (gray points in Figure \ref{fig;M_Re})
are included in the sample, such a trend is not changed.
Some other processes other than secular processes are needed for these less massive HAEs to evolve into local star-forming/quiescent galaxies.
We also find two massive, compact HAEs, and we will discuss these interesting objects in detail in Section \ref{sec;progenitor_quiescent}.

\section{Result}

\subsection{Dusty star-forming clump}
\label{sec;clump_migration}

A lot of HAEs at $z>2$ have kilo-parsec scale clumps as shown in Figure \ref{fig;clumpy_example}.
The fate of these clumps is an important issue and a matter of hot debate in relation to the bulge formation of galaxies.
In the numerical simulations, clumps formed in rotational disks can migrate toward galaxy centers as a result of their mutual interactions and of dynamical friction against the host disk, and coalesce into central young bulges \citep{2010MNRAS.404.2151C,2012MNRAS.422.1902I}.
This is a very efficient process to carry a large amount of gas from the galactic disks to the bulge components.
On the other hand, the momentum-driven galactic winds due to massive stars and supernova can disrupt giant clumps with $M_\mathrm{clump}=10^{8-9}M_\odot$ before they migrate towards galaxy centers \citep{2012ApJ...745...11G}.
\cite{2011ApJ...733..101G} and \cite{2012ApJ...752..111N} find the empirical evidence for gas outflows originating in massive luminous clumps in $z\sim2$ disks by deep AO-assisted integral field spectroscopic observations.
Whether the scenario for clump-origin bulge formation is viable depends sensitively on the longevity of clumps.
If the clumps contain a large amount of gas, they would survive and exhibit an age gradient as a function of distance from galaxy centers.
The \ha equivalent width is relatively insensitive to dust extinction if the line and continuum emissions
both originate in the same regions, and is thus the most useful measure of variation in stellar ages. 
%Note, however, that \ha line emission tends to be more attenuated than continuum light just beneath the line \citep{2005MNRAS.358..363C}, as the former originated only from most dusty star-forming regions.
\cite{2011ApJ...739...45F} have measured the equivalent widths of \ha emission (EW$_{\mathrm{H}\alpha}$) for clumps in one massive galaxy at $z\sim2$ and find a correlation between EW$_{\mathrm{H}\alpha}$ and galactocentric distance.
This suggests that the clumps near the galactic center tend to be older than the outer clumps, supporting the clump migration event.
All the other previous studies of clumps have relied on the colors of galaxies and SED fitting with multi-wavelength photometries.
\cite{2012ApJ...753..114W} have performed a detailed analysis of spatially resolved SEDs for a complete sample of star-forming galaxies at $1.5<z<2.5$. They find the trend of having redder colors, older stellar ages, and stronger dust extinction in the clumps near galactic centers compared to the off-center clumps.
\cite{2012ApJ...757..120G} have also shown the same obvious radial gradients in color, age and dust extinction for the clumpy galaxies at $z=1.5$--2.0.

\begin{figure}
\begin{center}
\includegraphics[scale=1.0]{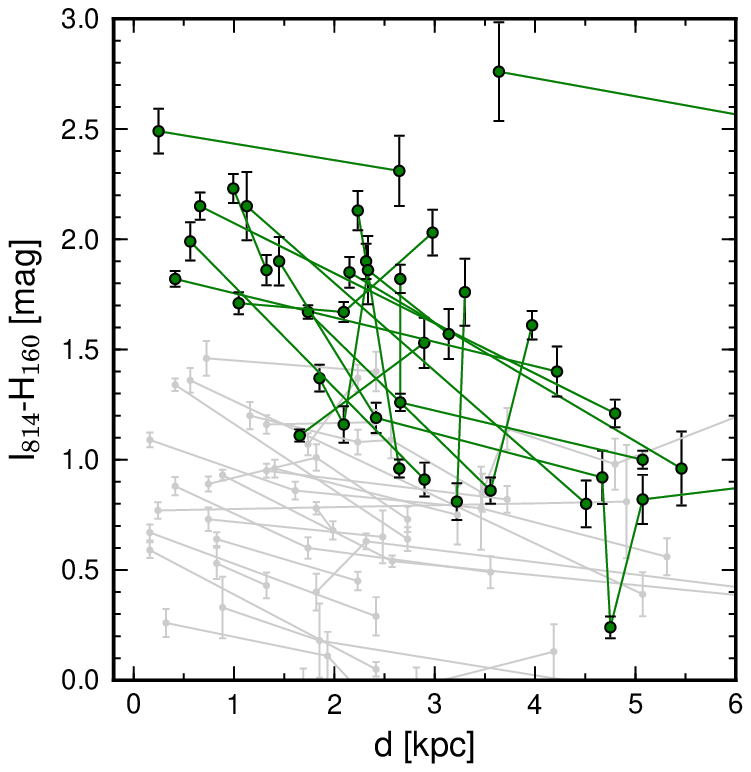}
\caption{The color gradients in $I_{814}-H_{160}$ of the clumps as a function of projected distance from a galactic center. The clumpy HAEs with a red clump of $I_{814}-H_{160}>1.5$ are presented. We also display other clumpy HAEs as gray symbols. \label{fig;clump_rvscolor}}
\end{center}
\end{figure}

In our sample, there are some clumpy HAEs with red clumps of $I_{814}-H_{160}>1.5$, which are often seen in the stellar mass range of $10^{10.5}~M_\odot<M_*<10^{10.8}~M_\odot$ (Figure \ref{fig;MS_clump}).
We focus on these HAEs with a red clump to investigate the clump properties and test the viability of the clump migration scenario.
Figure \ref{fig;clump_rvscolor} shows the radial gradient of clump colors across the host galaxies.
An aperture magnitude within a diameter of 0.36\arcsec (=2$\times$PSF size) is used to derive the color of a clump.
The stellar mass-weighted center is adopted to define a galactocentric distance.
We find that the clumps closer to the centers are redder compared to the off-center clumps in agreement with the previous studies.
The nuclear red clump seems to be a proto-bulge component, which would become an old bulge seen in local early-type galaxies.
%The issue here is which of the two (age or dust extinction) is responsible for the observed color variations among the clumps.
%It is challenging to give a clear-cut answer to this issue even with the current high-resolution, multi-wavelength data.

\begin{figure}
\begin{center}
\includegraphics[scale=1.0]{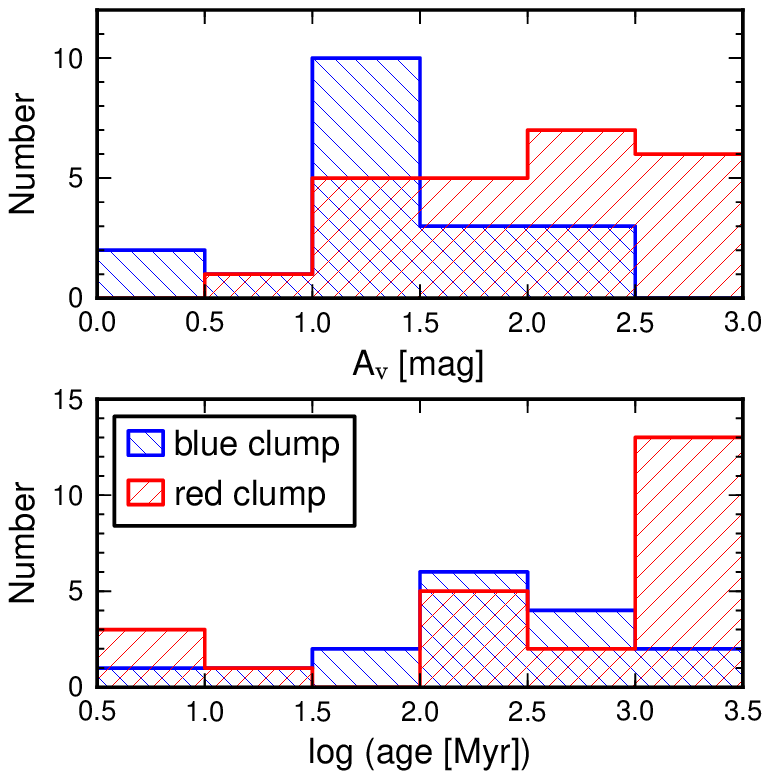}
\caption{Histograms of dust extinction and age of clumps divided according to color: $I_{814}-H_{160}>1.5$ (red hatched) and $I_{814}-H_{160}<1.5$ (blue hatched). Only HAEs with a red clump are used.\label{fig;hist_av}}
\end{center}
\end{figure}

Some HAEs with a red clump are detected in the public MIPS 24-micron image (SpUDS; PI: J. Dunlop).
Because the PSF size of the MIPS 24$\mu$m image is $\sim6$\arcsec, we cannot spatially resolve the infrared emission within galaxies for most of the HAEs, nor resolve the structures from clump to clump.
The five HAEs presented in Figure \ref{fig;clumpy_example} each have a blue clump of $I_{814}-H_{160}\simeq1.0$ as well as a red clump.
It is worth stressing that these objects are bright in the infrared emission, which indicates that a dusty starburst is occurring somewhere within them.
%The dust-obscured SFRs of overall galaxies are actually calculated to SFR$_\mathrm{IR}>200~M_\odot$yr$^{-1}$ from MIPS 24$\mu$m luminosities, assuming the template SEDs of \citep{2001ApJ...556..562C}.
If the blue clumps contain a large amount of dust, they should appear much redder due to dust extinction.
To pin down where the dusty star-forming regions are located within the galaxies,
we fit the SEDs of individual clumps with the stellar population synthesis model of \citep{2003MNRAS.344.1000B}  by using hyperz \citep{2000A&A...363..476B}.
Since only four broad-band photometries ($V_{606}, I_{814}, J_{125}, H_{160}$) are used for the fitting, constant star-formation histories and solar metallicity are assumed to estimate the amount of dust extinction and the luminosity-weighted age of the clumps.
In Figure \ref{fig;hist_av}, we plot histograms of dust extinction and luminosity-weighted age of the clumps.
We divide the clumps into red and blue clumps at $I_{814}-H_{160}=1.5$.
It is clear from the distributions that the red clumps are dustier and older than the blue clumps.
The Kolmogorov-Smirnov (K-S) tests suggest that the null hypothesis, that they are drawn from the same distribution, is proven to be false at the significance level of 1.0\% both in A$_v$ and age.
Therefore it is likely that dusty starburst activities are concentrated in the red clumps towards galaxy centers rather than in the blue clumps or in the inter-clump regions.

\begin{figure}
\begin{center}
\includegraphics[scale=1.0]{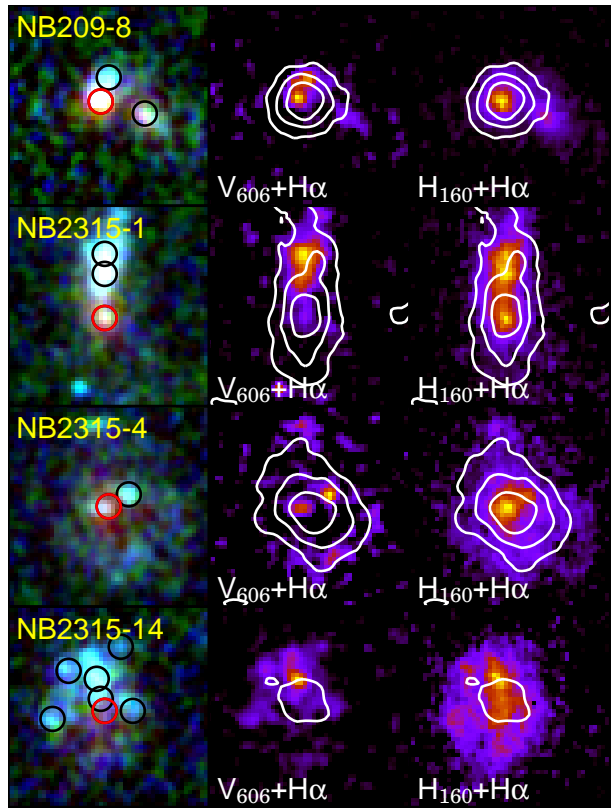}
\caption{Four HAEs with resolved dusty star-forming clumps. From left to right, three-color $V_{606}I_{814}H_{160}$, $V_{606}$-band images and $H_{160}$-band images are presented. Contours display the \ha flux density maps derived from NB (MOIRCS) and $K$-band (WFCAM; \citealt{2007MNRAS.379.1599L}) images whose PSF sizes are matched to $\sim0.7$\arcsec. Red and black circles in left panels indicate the positions of the reddest clump nearest to the galactic center and other bluer clumps, respectively.\label{fig;ha_map}}
\end{center}
\end{figure}

As an alternative approach, the \ha flux maps are created from the NB (line+continuum)
and $K$-band (continuum) images of the clumpy HAEs.
Since the \ha flux extends over entire galaxies in most cases, we cannot identify the internal star-forming regions from which strong \ha emission is actually radiated.
However, there are a few cases where we can resolve clumps even in the seeing limited \ha images. 
In Figure \ref{fig;ha_map}, we display the HST images of four such clumpy HAEs, which can be spatially resolved in the \ha maps.
If only the rest-frame UV luminosities are used as SFR indicators, 
the blue clumps would be seen as major contributors of the intense star formation within these galaxies.
However, the peak of \ha distribution is clearly located at or near the red clump rather than at the blue clumps.
Although \ha luminosities are sensitive to massive stars compared to UV, the two SFR indicators provide consistent results for old stellar population systems with age $>$ 100 Myr \citep{2013arXiv1310.5702W}.
Since the red clumps are older (Figure \ref{fig;hist_av}), the high flux ratios of \ha to UV is a robust evidence for dusty star formation in the red clumps.
\ha emission line is much less attenuated by dust compared to UV continuum emission, and penetrates through the dusty star-forming regions and comes out. 
If we can correct for the differential dust extinction among clumps, the intrinsic \ha line strength in the red clumps would be significantly larger and the star-formation map would become quite different from the UV luminosity map.
It should be noted that AGNs at the galactic centers may be contributing to MIPS 24$\mu$m fluxes and \ha flux densities.
At least for SXDF-NB209-6,7 and 11, however, the line ratios of H$\alpha$/[N~{\sc ii}] indicate that they are more like star-formation dominated galaxies rather than AGN-dominated ones (Tadaki et al., in preparation).
%We need additional spectroscopic follow-up observations to quantify AGN contributions for the rest of the objects.

%{2013arXiv1310.5702W}

Our results suggest that some HAEs at $z>2$ have a dusty star-forming proto-bulge component.
How is such a dusty starburst triggered at the center of the galaxies?
What is a feeding process to a proto-bulge component?
Although a major merger is a viable process to induce a nucleated starburst, 
other processes should be required in the case of normal star-forming galaxies with ordinary rotational disks.
Such star-forming clumps embedded in the disks are thought to be produced by gravitational instabilities of gas-rich disks \citep{2011ApJ...733..101G}.
However, clumps are seldom formed at the galactic center due to large velocity dispersion there.
The clump migration is a more preferable process than galaxy mergers, which can transport a large amount of gas from the rotational disks to the galactic center.
If gas rich clumps migrate to a galaxy center, a starburst would be induced at the center by a collision between the clumps in a manner similar to a major merger \citep{1996ApJ...471..115B}.

The gas fueling to the galactic center by clump migration is also an important process in view of the co-evolution of galaxies and supermassive black holes \citep{2011ApJ...741L..33B}.
Dusty star formation at the galaxy center is thought to accelerate the black-hole growth due to strong dissipation of gas and its further accretion towards the center.
This scenario is preferable therefore to account for the tight relationship between bulge and supermassive black hole masses \citep{2005Natur.433..604D}.

\subsection{Massive compact, star-forming galaxies}
\label{sec;progenitor_quiescent}

\begin{figure}
\begin{center}
\includegraphics[scale=1.0]{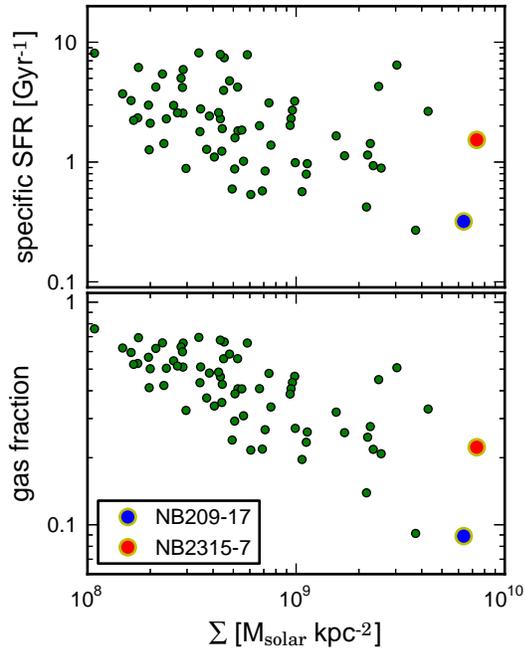}
\caption{Specific SFRs (top) and gas fractions (bottom) as a function of stellar surface densities ($\Sigma$=$M_*/2\pi r_e^{2}$) for the HAEs at $z=2.2$ and $z=2.5$. Symbols are the same as in Figure \ref{fig;M_Re}. The gas fraction is defined as $M_\mathrm{gas}/(M_\mathrm{gas}+M_\mathrm{star}$). \label{fig;sigmaM_sSFR}}
\end{center}
\end{figure}

Recent observational studies have shown that massive quiescent galaxies at $z\sim2$ tend to be extremely compact compared to local ones with the same stellar mass \citep[e.g.,][]{2008ApJ...677L...5V}. 
This suggests that the structures of compact quiescent galaxies have undergone a rapid evolution since $z\sim2$ \citep{2007MNRAS.382..109T}.
However, the progenitors and the formation processes of such compact quiescent galaxies have been much less investigated.
\cite{2013ApJ...765..104B} have studied both the number densities of compact quiescent galaxies (``red nuggets'') and compact star-forming galaxies (``blue nuggets'') at $z=1.3$--3.0, 
and find that blue nuggets can fade to red nuggets with the starburst lifetime of $\sim0.8$ Gyr. 
In our \ha emitter sample, we find two massive, compact star-forming galaxies with M$_*>10^{11}$ M$_\odot$ and $r_\mathrm{e}<2$ kpc (SXDF-NB209-17 and SXDF-NB2315-7 in Figure \ref{fig;M_Re}).
Their surface stellar mass densities are as high as those of red nuggets at $z=1.5$--2.0 \citep{2012ApJ...746..162N}.
Moreover, the two compact HAEs show high S$\acute{\mathrm{e}}$rsic indices of $n\geq2$, which indicate relaxed bulge-like morphologies.
They can directly turn into compact quiescent galaxies at high-redshift just by stopping their star-formation activities without changing the sizes or structures.
%We also find another \ha emitter with a similar stellar surface density, but its structure is somewhat extended ($r_\mathrm{e}\sim3$ kpc) with a gaussian-like profile ($n=0.5$).
%Therefore, we treat only the two HAEs (SXDF-NB209-17 and SXDF-NB2315-17) as star-forming nuggets in this paper.
Note that SXDF-NB209-17 has been spectroscopically confirmed to be located at $z = 2.182$ (Tadaki et al., in preparation).

The star-formation activities and stellar populations substantially differ between the two compact HAEs.
Figure \ref{fig;sigmaM_sSFR} shows the relation between stellar surface density (defined as $M_*/2\pi r_e^{2}$) and specific SFR (=SFR/$M_*$) for the HAEs.
SXDF-NB209-17 is found to have significantly suppressed specific SFRs compared to the other extended HAEs. 
This may suggest that the star-forming activity in the compact galaxy is just beginning to be quenched.
In contrast, the star-formation activity of SXDF-NB2315-7 is still high despite its large stellar mass of $M_*>10^{11}M_\odot$.
Next, we evaluate the gas mass of HAEs from $\Sigma_\mathrm{SFR}$ with the assumption of the Kennicutt-Schmidt relation, $\Sigma_\mathrm{SFR}=(2.5\times10^{-4})\Sigma_\mathrm{gas}^{\mathrm{1.4}}$ \citep{1998ARA&A..36..189K}.
While most of the HAEs show high gas fractions of $\sim0.5$, consistent with the direct measurements of gas fraction in star-forming galaxies at $z\sim2$ based on the CO observations \citep{2010Natur.463..781T,2010ApJ...713..686D}, 
the gas fractions of the two nuggets are clearly much lower, especially for SXDF-NB209-17.
This further supports the idea that they are in the transitional phase from star-forming to quiescent galaxies.

\begin{figure}
\begin{center}
\includegraphics[scale=1.0]{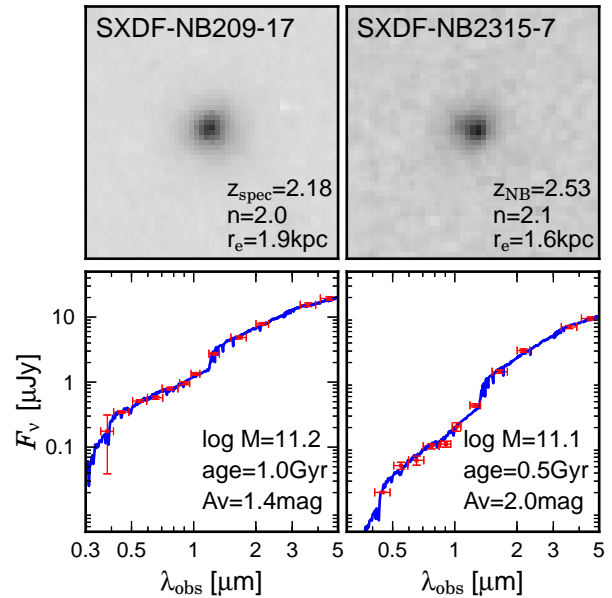}
\caption{$H_{160}$-band images (top) and SEDs (bottom) of two massive compact star-forming galaxies SXDF-NB209-17 (left) and NB2315-7 (right). 
Red symbols and blue spectra indicate the observed SEDs and the best-fit SED models by Bruzual \& Charlot (2003), respectively. 
\label{fig;compact_SED}}
\end{center}
\end{figure}

Also, the SED from UV to near-infrared indicates an old stellar population with the age of 1.0 Gyr in SXDF-NB209-17, while SXDF-NB2315-7 is likely to be a young dusty star-bursting galaxy with the age of 0.5 Gyr and dust extinction of 2.0 mag in $A_V$ (Figure \ref{fig;compact_SED}).
These results suggest that both of them are good candidates for the progenitors of red nuggets at $z=1.5-2.0$, but they are in the midst of the different evolutionary phases on the way to quiescent galaxies.
While SXDF-NB2315-7 is in the starburst phase, SXDF-NB209-17 is probably a similar population to the young quiescent (post-starburst) galaxies reported by \cite{2012ApJ...745..179W}.
Such a compact star-forming phase appears only at $z>1$ and it can be the direct channel to form red nuggets seen at similar redshifts.

\section{Formation of massive quiescent galaxies}

\begin{figure}
\begin{center}
\includegraphics[scale=1.0]{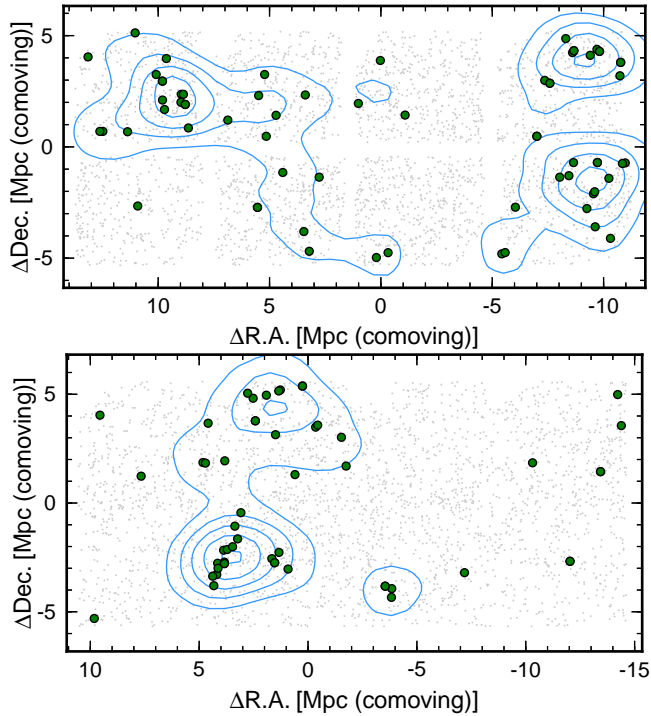}
\caption{The spatial distributions of the HAEs at $z = 2.2$ (top) and $z = 2.5$ (bottom). Green filled circles indicate the HAEs, while small gray dots show all the objects detected in the NB images. The contours present the smoothed local number density of the HAEs ($\Sigma_\mathrm{5th}=$2, 4, 6, 8, 10 Mpc$^{-2}$).
\label{fig;distribution}}
\end{center}
\end{figure}

We have presented two kinds of curious populations: galaxies with a nuclear dusty star-forming clump, and the very compact star-forming nuggets.
The fundamental questions are how they evolve afterwards and what their descendant galaxies are in the present Universe.
The constant number density method is a useful approach with which we can link high-redshift galaxies to the local ones \citep[e.g.,][]{2013ApJ...766...33L}.
As shown in Figure \ref{fig;MS_clump}, the completeness limit of our survey almost reaches down to $M_*>10^{10}~M_\odot$.
The stellar masses of the HAEs with red clumps and the star-forming nuggets are $M_*>10^{10.5}~M_\odot$ (Figures \ref{fig;MS_clump} and \ref{fig;compact_SED}), which corresponds to the number density of $n=(5$--$6)\times10^{-4}$ Mpc$^{-3}$ \citep{2013ApJ...766...15P}.
At the constant number density, their descendants would be massive galaxies with $M_*>10^{11}~M_\odot$ at $z\sim0$, and dominated by red, quiescent galaxies \citep{2004ApJ...600..681B}.
Moreover, our HAEs are inhomogeneously distributed and constitute large-scale structures at $z>2$ (Figure \ref{fig;distribution}).
The galaxies existing in such overdense regions at high redshifts are most likely the progenitors of massive, quiescent galaxies like giant ellipticals in the local clusters/groups, rather than late-type galaxies that dominate in less dense environments.
In this section, we will discuss the evolutionary paths of our sample of star-forming galaxies at high redshifts to the massive quiescent galaxies at later epochs.

\begin{figure*}
\begin{center}
\includegraphics[width=160mm]{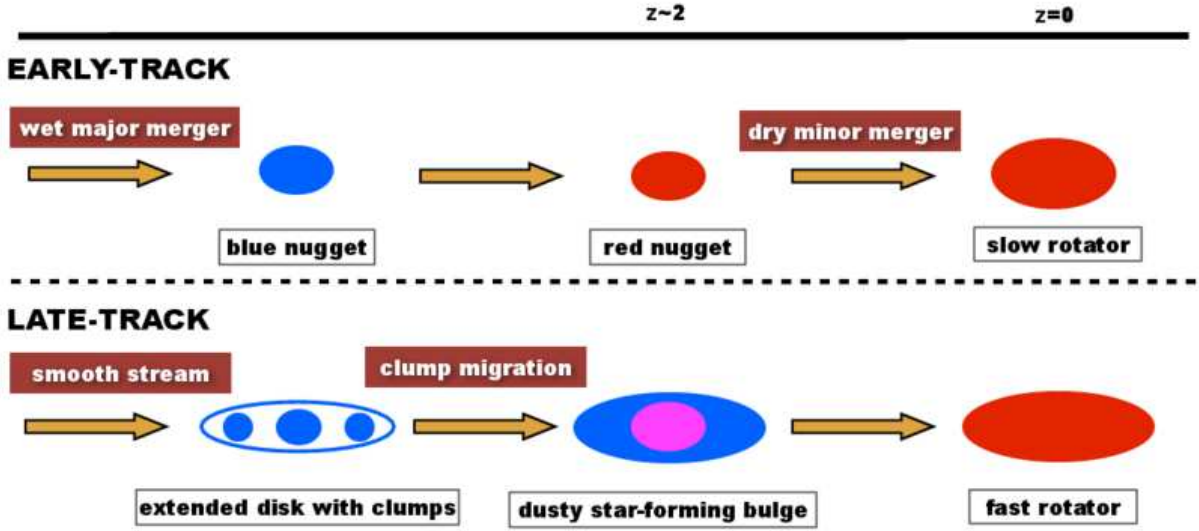}
\caption{Schematic pictures of the proposed evolutionary tracks from high-redshift star-forming galaxies to local massive, quiescent galaxies. Blue, red and magenta present star-forming, quiescent and dusty star-forming components, respectively. \label{fig;track}}
\end{center}
\end{figure*}

We present the schematic chart of the two kinds of evolutionary tracks from high-redshift galaxies to local quiescent ones in Figure \ref{fig;track}.
%The merger paradigm accounts for both the formation of massive compact star-forming galaxies and the transformation of galaxy morphologies.
In the numerical simulations, a gas-rich major merger at high redshifts can produce a massive, compact system \citep{2009ApJ...700..799W,2011ApJ...730....4B}.
Our two star-forming nuggets are likely to be in the phase of violent starburst or shortly after that.
They would subsequently quench star formation (and become red nuggets), and further evolve from red nuggets into large quiescent galaxies through a number of dry minor mergers.
By forepast dissipational processes such as a major merger, they would be dispersion-dominated objects, and be observed as slow rotators, early-type galaxies in the present Universe \citep{2011MNRAS.414..888E}. They would show little or no rotation.
This evolutionary path would be most efficient at high redshifts because almost all the star-forming galaxies are no longer compact at $z < 1$, and therefore this path can be called $early\ track$ \citep{2013ApJ...765..104B}.

%While the number density of our star-forming nuggets is $n_\mathrm{cHAEs}=8\times10^{-5}$ Mpc$^{-3}$ at $z>2$, massive quiescent galaxies are more frequently seen at $z=1.7$ ($n_\mathrm{QG}=(2-3)\times10^{-4}$ Mpc$^{-3}$; \citealt{2011ApJ...739...24B}).
Though the short duty cycle of star-forming nuggets can explain the evolution of number density of red nuggets \citep{2013ApJ...765..104B},
the number density of massive quiescent galaxies increases by a factor of two between $z=1$ and $z=0$ \citep{2011ApJ...739...24B}, at which the nuggets would be very rare.
This means that there must be another evolutionary path to form quiescent galaxies from the majority of star-forming galaxies. %%ZZZ- needs a number density argument of the early/late-track galaxies here.
The size evolution during the star-forming phase before they quench star formation is also quite important as well as the later growth of their sizes by minor mergers. 
We find a lot of star-forming galaxies with extended disks at $z>2$ (Figure \ref{fig;M_Re}).
They are expected to evolve into more massive galaxies just following the local mass--size relation, unless some external processes such as mergers happen to occur.
The feeding process through cold gas accretions would calm down at $z<1.5$, because even typical dark matter halos become as massive as $M_\mathrm{halo}\sim10^{11.5-12.0} M_\odot$ and the gas within the halos is heated up by a virial shock \citep{2006MNRAS.368....2D}.
The remaining cold gas within galaxies is gradually consumed and star-formation activities would be significantly decreased due to the inefficient supply of cold gas.
Therefore, the HAEs with extended disks would quench their star formation eventually at later times, and directly evolve into large quiescent galaxies that are more frequently present at $z\sim1$, just by the consumption of gas. This path for the latecomers of quiescent galaxies can be called $late\ track$ \citep{2013ApJ...765..104B}.

Massive, large quiescent galaxies through the late track should be observed as fast rotator early type galaxies at $z\simeq0$, provided that their progenitors do not undergo dissipational processes \citep{2013MNRAS.432.1862C}.
However, the formation process of bulge components is still unknown in the late track.
It needs the transformation of structures as well as the size evolution.
We show that a lot of HAEs have clumpy morphologies. We also find that nuclear dusty star-forming clumps in the HAEs with $M_*\sim10^{10.5}~M_\odot$ are the expected sites of bulge formation.
A gas supply through giant clumps could fully explain such a dusty star-forming proto-bulge although we do not have any definitive evidence for it yet.
Since a galactic wind at each clump may disrupt the clumps before they migrate to the center, such feedback process is also critical to destine the subsequent evolution of the clumps \citep{2012MNRAS.427..968H}.
To validate the scenario of clump migration, we have to know whether the clumps
contain gas that is massive enough to be left over after a rapid consumption by high star-forming activity and a subsequent strong stellar feedback.
High-resolution observations with JVLA and/or ALMA provide us with valuable information about the amount of molecular gas within clumps.
Our discovery of dusty star-forming clumps is the first step to verifying the clump migration scenario to form galactic bulges.

\section{Summary}

%We note that most of the rejected objects whose structural parameters are not obtained due to
%bad fitting results, show faint, extended structures (see also Figure \ref{fig;lowG}).

In this paper, we present and discuss the morphological properties of \ha selected star-forming galaxies at $z>2$ in SXDF-UDS-CANDELS field.
Using the high-resolution images that trace the rest-frame UV and optical light,
many kilo-parsec scale clumps are identified within galaxies by a semi-automatic method.
We find that at least 41\% of our sample show clumpy structures and the fraction of clumpy galaxy peaks in $M_*\sim10^{10.5}M_\odot$.
We find no significant difference in the distribution on the offset from the main sequence between clumpy and non-clumpy galaxies, however.
%If a higher spatial resolution image resolves smaller clumps, we might see clumpy galaxies more frequently.

For HAEs containing a red clump, it is found that the clump closest to the galaxy center is redder compared to the off-center clumps in the disks. 
The presence of infrared emission and the \ha flux density maps of four of them suggest that dusty star formation is most likely occurring in the nuclear red clump.
Since it is unlikely that a clump is directly formed from gravitational instabilities in a galaxy center with such large velocity dispersion,
one needs a feeding process to drive a large amount of gas to the center.
The clump migration to the galactic center and the subsequent starburst there seem to be a promising mechanism in galactic bulge formation at $z\sim2$.

Also, we obtain the structural parameters (effective radius and S$\acute{\mathrm{e}}$rsic index) for 70 objects with $M_*>4\times10^9M_\odot$.
Seventy-four percent of the HAEs at $z>2$ seem to have disk-like morphologies according to their S$\acute{\mathrm{e}}$rsic indices.
The similarity of the size--mass relation for the majority of the star-forming galaxies at $z>2$ to the local relation may suggest that many galaxies have already obtained disks as extended as the local ones.
However, it is notable that we observe two massive compact star-forming galaxies at $z>2$.
One has slightly smaller specific SFR than others, suggesting that star formation is just being quenched.
In contrast, the other is an actively star-forming galaxy with a high specific SFR.
Given their high stellar surface densities, these compact HAEs are likely to be the direct progenitors of compact quiescent galaxies at $z=1.5-2.0$.
Their S$\acute{\mathrm{e}}$rsic indices of $n\sim2$ also support bulge-dominated morphologies.

They would then later evolve by minor mergers, particularly in terms of size, into massive quiescent galaxies by the present day. 
If these massive compact HAEs at $z>2$ are in the post-starburst or in the starburst phases due to gas-rich major mergers,
they would be observed as slow rotator, early-type galaxies in the present-day Universe.

On the other hand, to account for the increase in the number density of quiescent galaxies from $z\sim1$ to $z=0$, 
we consider another evolutionary path from high-redshift star-forming galaxies to local quiescent galaxies; these star-forming galaxies continuously grow both in size and in mass along the size--mass relation and develop large disks, and at some point they eventually stop their star-formation activity and directly evolve into large quiescent galaxies as latecomers.
In our sample, we actually find a lot of massive HAEs following the local size--mass relation.
The clump migration would be the preferred mechanism in this scenario to grow bulge component.
They would end up with fast rotator, early-type galaxies at $z=0$.

Our unique, unbiased sample of star-forming galaxies at $z>2$ and their morphological properties have revealed that these two evolutionary paths can successfully provide us with reasonable understanding of the overall evolution of massive quiescent galaxies from $z=2$ to $z=0$.

%\end{enumerate}

In this work, we treat a clumpy galaxy as a single disk galaxy.
However, it also resembles a merging system, and we are not sure of such a hypothesis with the current stellar continuum map alone.
We need to investigate the kinematics of the HAEs by IFU spectroscopy at near-infrared with AO as well as CO spectroscopy in radio to reveal their physical states based on both internal velocity and spatial structures.
We also note that our analyses do not fully consider the effects of AGNs since spectroscopic follow-up observations have been conducted only for
a part of the HAEs.
By performing a near-infrared spectroscopy on a complete, statistical sample of our HAEs,
we must be able to understand how much AGN feedback actually contributes to the
quenching of star-formation in massive galaxies at the peak epoch of their formation.

\

\

This paper is based on data collected at Subaru Telescope, which is operated by the National Astronomical Observatory of Japan. 
We thank the Subaru telescope staff for their help in the observation. 
We thank the anonymous referee who gave us many useful comments, which improved the paper.
K.T. would also like to thank Dr. Kazuhiro Shimasaku, Professor Masashi Chiba, Dr. Nobunari Kashikawa, Dr. Kentaro Motohara, Dr. Masami Ouchi, and Professor Masanori Iye for useful discussions and comments.
K.T. and Y.K. acknowledge the support from the Japan Society for the Promotion of Science (JSPS) through JSPS research fellowships for young scientists.  
T.K. acknowledges the financial support in part by a Grant-in-Aid for the Scientific Research (Nos.\, 18684004, 21340045, and 24244015) by the Japanese Ministry of Education, Culture, Sports, Science and Technology.

\bibliographystyle{apj}
\bibliography{tadaki_2013}

\end{document}